\documentclass{PoS}

\usepackage{amsmath}
\usepackage{amssymb}
\usepackage{amstext}
\usepackage{amsfonts}
\usepackage{subfigure}
\usepackage{multirow}

\newcommand{\ba}{\begin{eqnarray}}
\newcommand{\ea}{\end{eqnarray}}
\newcommand{\be}{\begin{equation}}
\newcommand{\ee}{\end{equation}}
\newcommand{\bd}{\begin{displaymath}}
\newcommand{\ed}{\end{displaymath}}
\newcommand{\een}{\nonumber\end{equation}}
\newcommand{\bea}{\begin{eqnarray}}
\newcommand{\eean}{\nonumber\end{eqnarray}}
\newcommand{\eea}{\end{eqnarray}}

\def\eq#1{Eq.~(\ref{#1})}

\def\fig#1{Fig.~\ref{#1}}

\newcommand{\gap}{\hspace{10pt}}

\newcommand{\mev}{\mathrm{MeV}}

\newcommand{\fm}{~\mathrm{fm}}

\newcommand{\beq}{\begin{equation}}   
\newcommand{\eeq}{\end{equation}}   
\newcommand{\beqn}{\begin{eqnarray}}  
\newcommand{\eeqn}{\end{eqnarray}}

\def\mcO{{\mathcal O}}

\def\la{\langle}
\def\ra{\rangle}

\newcommand{\old}[1]{}

\title{\vspace*{-2cm}{\large \normalfont \hfill DESY 12-213, SFB/CPP-12-89} \\\vspace*{1cm}
Sigma terms and strangeness content of the nucleon with $Nf=2+1+1$ twisted mass fermions\vspace*{0.3cm}
}
\ShortTitle{Excited state contaminations and $\sigma$-terms of the nucleon}

\author{C. Alexandrou$^{ab}$, M. Constantinou$\,^b$,~S. Dinter$\,^c$, \speaker{V. Drach}$\,^c$, K. Hadjiyiannakou$^b$, K. Jansen$\,^c$ , G. Koutsou$^a$,~A. Strelchenko$\,^b$ and A.~Vaquero$^a$ \\
\llap{$^a$} Computation-based Science and Technology Research Center (CaSToRC), The Cyprus Institute, \\
20 Constantinou Kavafi Street Nicosia 2121, Cyprus \\
\llap{$^b$} Departament of Physics, University of Cyprus, P.O. Box 20537, 1678 Nicosia, Cyprus \\
\llap{$^c$}{NIC, DESY Zeuthen, Platanenallee 6, D-15738 Zeuthen, Germany\\}

E-mail: \email{vincent.drach@desy.de}

}

\abstract{We investigate excited state contaminations in a direct computation of the nucleon $\sigma$-terms. 
This is an important source of systematic effects that needs to be controlled besides the 
light quark mass dependence and lattice artefacts. We use maximally twisted mass fermions with 
dynamical light ($u$,$d$), strange and charm degrees of freedom. Employing an efficient 
stochastic evaluation of the disconnected contribution available for twisted mass fermions, 
we show that the effect of excited states is large in particular for the strange 
$\sigma$-terms, where it can be as big as $O(\gtrsim 40\%$). This leads to the 
unfortunate conclusion that even with a source-sink 
separation of 
$\sim 1.5 \fm$ and a good statistical accuracy it is not clear, whether  
excited state effects are under control for this quantity.}

\FullConference{The 30 International Symposium on Lattice Field Theory - Lattice
 2012,\\
 June 24-29, 2012\\
 Cairns, Australia}

\begin{document}

\section{Introduction}

The various evidences for the existence of dark matter have led to the development of 
experiments dedicated to detect dark matter directly. The detection relies on the 
measurements of the recoil of atoms hit by a dark matter candidate. One popular class of 
dark matter models involve an interaction between a WIMP and a Nucleon mediated by a Higgs exchange.  
Therefore, the scalar content of the nucleon is a fundamental ingredient in the WIMP-Nucleon 
cross section. In this way, the uncertainties of the scalar content translate directly  
into the accuracy of the constraints on beyond the standard model physics. 
Since the coupling of the Higgs to quarks is proportional to the quark masses, 
it is important to know how large scalar matrix elements of the nucleon are, 
in particular for the strange and charm quarks.  

One common way to write the parameters entering the relevant cross section are the 
so-called sigma-terms of the nucleon:
\be\label{eq:sigma_terms}
\sigma_{\pi N} \equiv m \la N|  \bar{u} u+  \bar{d} d |N \ra \gap\textmd{and}\gap 
\sigma_s  \equiv  m_s\la N|  \bar{s}s |N \ra\; ,
\ee
where $m$ denotes the light quark mass and $m_s$ the strange quark mass. 
To quantify the scalar strangeness content of the nucleon 
a parameter $y_N$ is introduced,

\be\label{eq:y_para}
y_N \equiv  \frac{ 2\la N| \bar{s} s |N \ra }{ \la N| \bar{u} u+ \bar{d} d |N \ra},
\ee
which can be also related to the sigma terms of the nucleon in eq.~(\ref{eq:sigma_terms}).

The direct computation of the above matrix elements is known to be challenging on the lattice for 
several reasons. First, it involves the computation of "singlet"  or "disconnected'' diagrams 
that are very noisy. Second, discretisations that break chiral symmetry generally 
suffer from a mixing under renormalization between the light and strange sector, 
which is difficult to treat in a fully non-perturbative way.  

However, as has been shown in \cite{Dinter:2012tt}, twisted mass fermions offer two 
advantages here: they provide both an efficient variance noise reduction for 
disconnected diagrams~\cite{Jansen:2008wv} and avoid the chirally violating 
contributions that are responsible for the mixing under renormalization.

A great effort has been spent developing techniques to estimate 
efficiently the relevant disconnected contribution 
(see for instance \cite{Alexandrou:2012py,Engelhardt:2012gd,Oksuzian:2012rzb,Bali:2012qs}). 
Also excited state contributions to nucleon matrix elements connected 
to deep inelastic scattering received a lot of attention 
during the past years \cite{Capitani:2012gj,Dinter:2011sg}. 
However, the determination 
of systematic effects and in particular the excited state contaminations of the nucleon $\sigma$-terms
are so far quite limited and it is the main goal of this contribution to investigate these excited state effects.

\section{Lattice Techniques}

In this study we used a single set of gauge configuration produced by the ETM collaboration.
We used a $N_f=2+1+1$ ensemble with a pion mass of $390 \mev$, a volume $V=32^3\times 64$ 
and a lattice spacing $a\approx 0.078\fm$. We refer to \cite{Baron:2010bv} for details 
on the gauge ensemble used in this work. In order to compute matrix elements involving 
strange quarks, we  work within a mixed action setup introducing an additional doublet 
of degenerate twisted mass quarks of mass $\mu_q$ in the valence sector.

The scalar matrix elements involved in \eq{eq:sigma_terms} can then be computed using the asymptotic 
behaviour of a suitable ratio of three and two-point functions defined as 
\be
\label{eq:ratio_def}
R_{O_q}(t_s, t_{\rm op}) = \frac{C^{O_q}_{\rm 3pts}(t_s,t_{\rm op})}{C_{\rm 2pts}(t_s) } = \la N | O_q | N \ra^{(\rm bare)} + \mcO( e^{-\Delta  t_{\rm op}}) +  \mcO( e^{-\Delta (t_s - t_{\rm op})} )\;,
\ee
where $O_q$ refers to the operator in which we are interested in, 
namely $O_l \equiv \bar{u}u + \bar{d}d$ and $O_s\equiv\bar{s}s$. In eq.~(\ref{eq:ratio_def}) 
$t_s$ refers to the source-sink separation and $t_{\rm op}$ to the source-operator separation. 
In addition, $\Delta$ represents the mass gap between the nucleon and its first excited state.  
From eq.~(\ref{eq:ratio_def}) it is clear that large times $t_{\rm op}$ and $t_s$ are 
needed to suppress the so-called excited state contributions. 
However, due to the exponential decrease of the signal-over-noise ratio at large times, it is 
numerically very expensive to obtain a good signal for increasing $t_s$ or $t_{\rm op}$. 

The nucleon states themselves are created using smeared interpolating fields, which have 
already 
been optimized themselves to suppress excited state contaminations in the two point function. 
Using the same gauge field ensemble and interpolating fields, it was shown in \cite{Dinter:2011sg} 
that the axial coupling of the nucleon, $g_A$, can be safely extracted with $t_s = 12a \sim 0.9 \fm$.

Instead of using the time dependence in $R_{O_q}$ of eq.~(\ref{eq:ratio_def}) and
looking for a plateau behaviour, another way to extract the 
desired matrix element is to consider the so-called summed ratio method. 
Integrating \eq{eq:ratio_def} over the time of insertion of the operator we 
are left we the following asymptotic behaviour:
\be\label{eq:sum_ratio_def}
P_{O_q}( t_s) = \sum_{t_{{\rm op}=0}}^{t_s} R_{O_q}(t_s, t_{\rm op})  = A +  \la N | O_q | N \ra^{(\rm bare)} t_{s}+ \mcO( e^{-\Delta  t_{s}}) \; .
\ee

For the precise expression of the operators  $O_q$, their multiplicative renormalization and 
our computational techniques we refer
the reader to \cite{Dinter:2012tt}. As presented in \cite{Alexandrou:2012py} part of the computation has been done making intensive use of a modified version of the QUDA library \cite{QUDA1,QUDA2}.

\section{Excited state contaminations: $\sigma$-terms}

\begin{figure}[h]
\begin{minipage}[ht]{7.5cm}
\includegraphics[height=6cm,width=7.5cm]{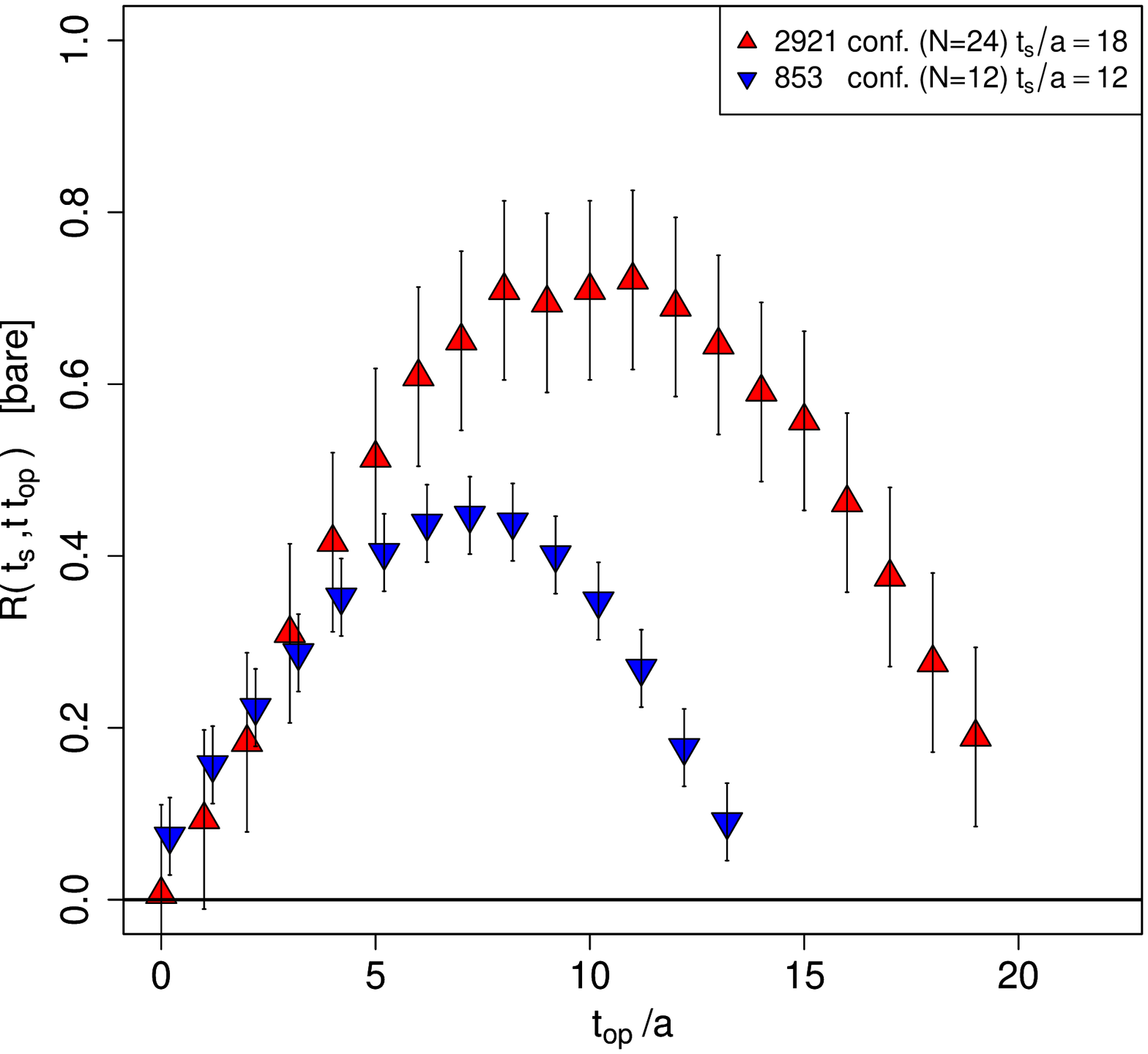}
\caption{The ratios $R_{O_s}(t_s=12a,t_{\rm op})$ and $R_{O_s}(t_s=18a,t_{\rm op})$ measured with 
a statistics of $\approx 800$ and $\approx 3000$ configurations as a function of 
$t_{\rm op}$ for fixed $t_s$.}\label{fig:plateau_strange}
\end{minipage}
\hspace{0.5cm}
\begin{minipage}[ht]{7.5cm}
\vspace{+0.45cm}
\includegraphics[height=6cm,width=7.5cm]{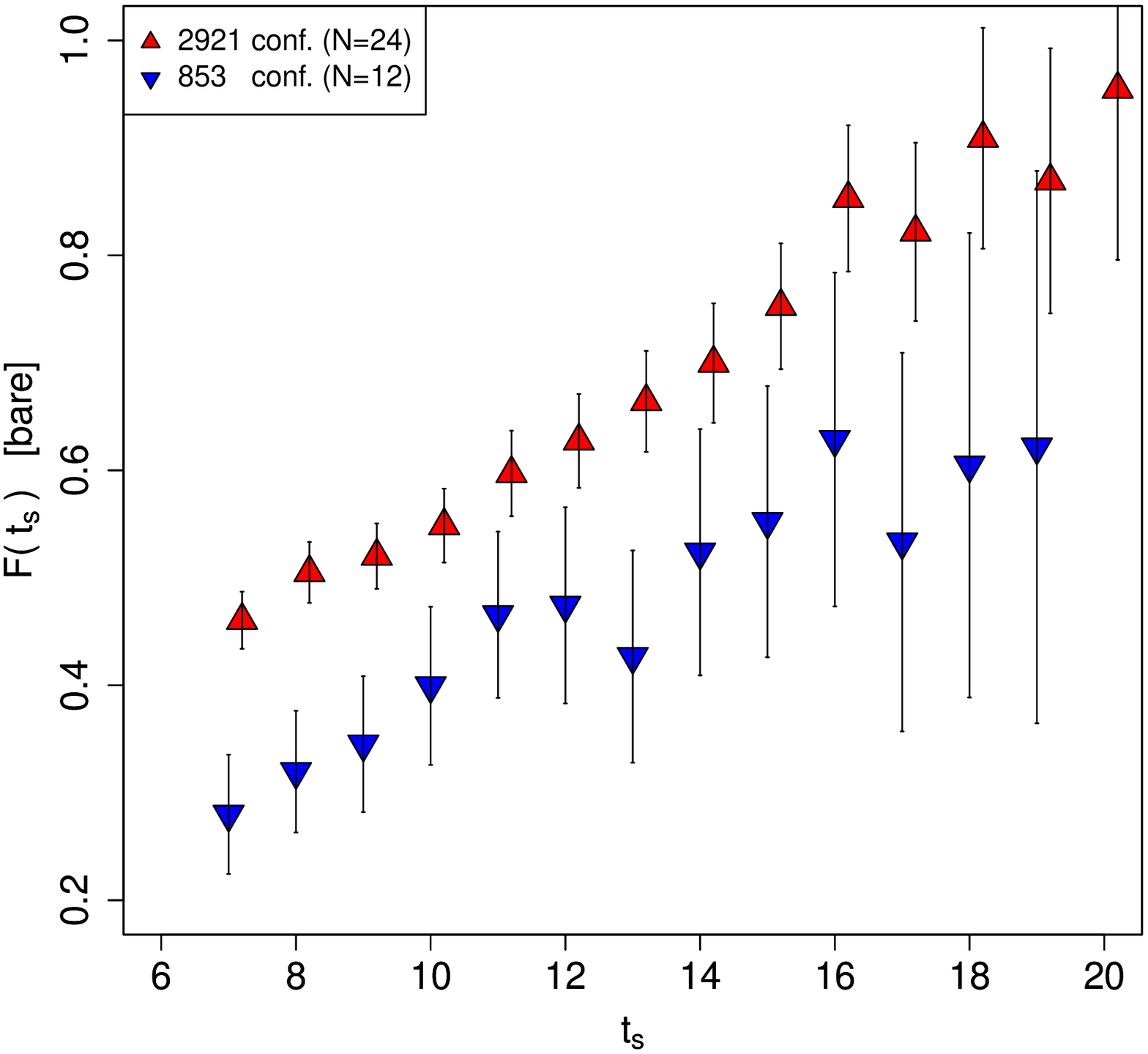}
\caption{Values for the plateaux in $t_{\rm op}$ of eq~(\protect\ref{eq:ratio_def}) as a function of $t_s$, for the two sets of measurements $A$ and $B$. The points have been shifted vertically  for better readability for the sample $B$.}\label{fig:fit_plateau_strange_small_vs_largestat}
\end{minipage}
\end{figure}
 
We first concentrate on the computation of $\la N | \bar{s}s | N \ra^{(\rm bare)}$. 
Note that once multiplied by the bare strange quark mass, we obtain 
the strange $\sigma-$term of the nucleon $\sigma_s$ defined in \eq{eq:sigma_terms}. 
Since $\la N | \bar{s}s | N \ra^{(\rm bare)}$ only involves 
disconnected diagrams, with our techniques we can  
freely change the source-sink separation $t_s$. 

In the following we will refer to a small statistics 
sample ($\approx 800$ configurations) as sample $A$ while a large statistics sample ($\approx 3000$ configurations) will be referred to as sample $B$\footnote{The exact number of configurations used is 853 and 2921, respectively.}. Note that we checked that the number of noise vector used to estimate the disconnected loops is large enough  to ensure that the statistical errors are dominated by gauge noise.

In \fig{fig:plateau_strange} we show the ratio $R_{O_s}$ obtained at a 
fixed value of $t_s$ as a function of $t_{\rm op}$.  We show the results for $t_s=12a$ on the sample $A$ and for $t_s=18a$ on the sample $B$.
In the graph we show constant fits of the ratio $R_{O_s}$ as function of $t_{\rm op}$ 
choosing a
fitting interval $t_{\rm op} \in [t_s/2-1,t_s/2+1]$ 
($t_{\rm op} \in [(t_s/2+\frac{1}{2})-2,(t_s/2-\frac{1}{2})+2]$ in case
$t_s$ is odd).  
Comparing samples $A$ and $B$, we observe an increase of the so fitted 
plateaux values of $\sim 40\%$. This indicates that source-sink 
separations of $ \sim 0.9 \fm$ are not large enough to have negligible systematic errors 
from excited states. 
It is important to note that although we have chosen a quite large  
source-sink separation for sample $B$ ($ t_s = 18a$) 
we reached a statistics giving  
comparable statistical error to the one of sample $A$ ($ t_s = 18a$). 
In order to obtain a better understanding of the   
systematic errors coming from the excited state contaminations, we analyze the 
strange quark content of the nucleon using both the ratio and and the summed ratio method.

As mentioned above, our fit interval in $t_{\rm op}$ for $R_{O_s}$ is always chosen 
symmetrically around $t_s/2$. Therefore, for each choice of $t_s$ 
we obtain different plateaux values providing in this way a function 
$F(t_s)$. Ideally, when effects of excited states are sufficiently suppressed, 
$F(t_s)$ reaches a plateau value which would correspond to the true
bare matrix element $\la N | \bar{s}s | N \ra^{(\rm bare)}$. 
We show in \fig{fig:fit_plateau_strange_small_vs_largestat} $F(t_s)$ as a function 
of the source-sink separation $t_s$ for the samples $A$ and $B$. For better
readability we   
shifted vertically 
the results obtained from sample $B$. Looking first only at sample $A$, 
it seems that $F(t_s)$ indeed reaches a plateau value for $t_s \gtrsim 11$ from which 
we would conclude that the asymptotic value has been found. 
However, going to sample $B$ which has a higher statistics, it becomes 
clear that $F(t_s)$ still increases beyond $t_s > 11a$ and may only 
reach a plateau like behaviour for $t_s \gtrsim 16$. 
This finding provides a serious warning. In fact, 
we cannot exclude that even with a 
larger statistics the required value of $t_s$ would be larger 
to observe a clear plateau. 
 
\begin{figure}[h]
\begin{minipage}[ht]{7.5cm}
\vspace{+0.65cm}
\includegraphics[height=6cm,width=7.5cm]{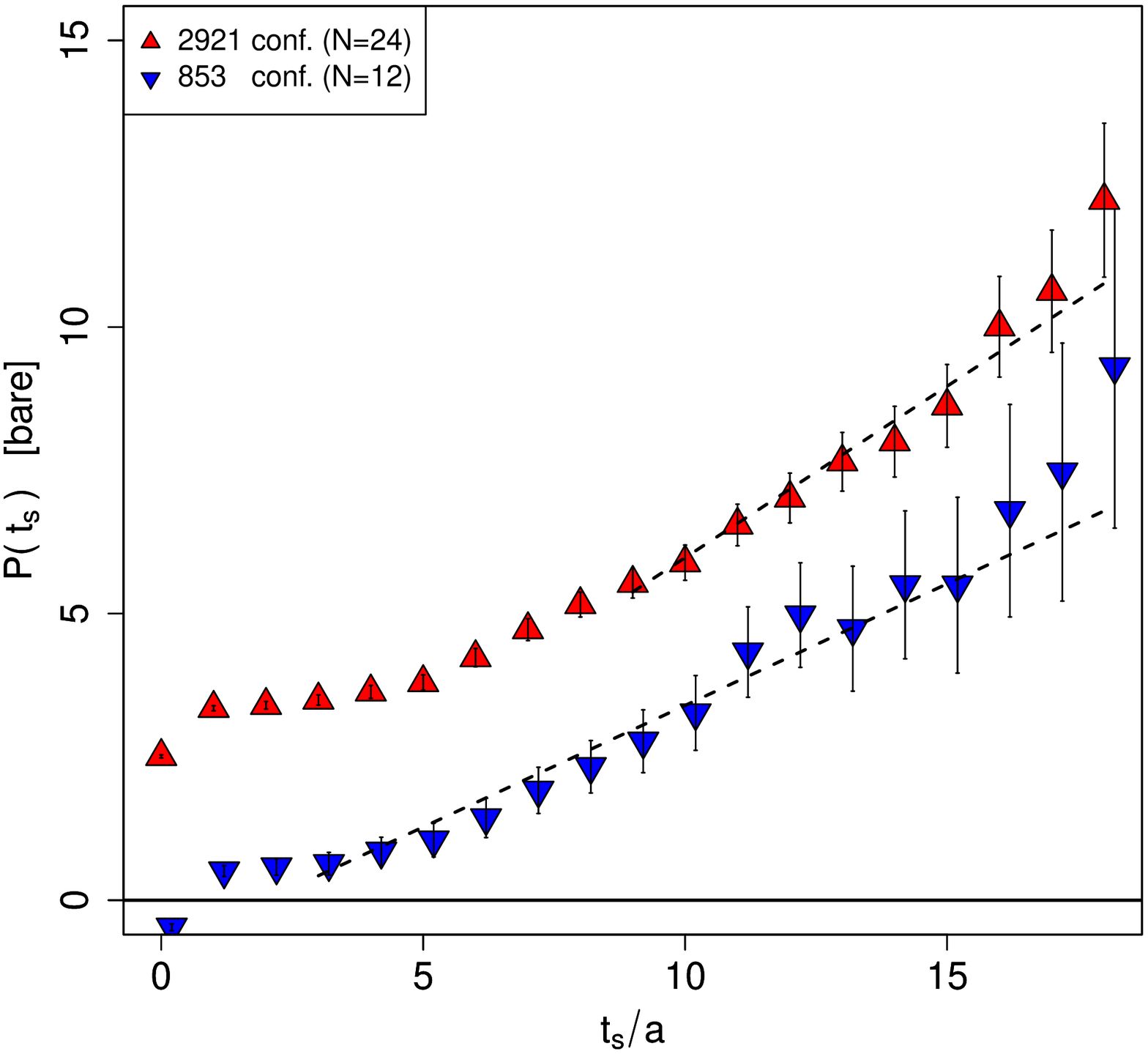}
\caption{Summed ratio $P_{O_s}$ as a function of the source-sink separation 
for the two samples $A$ and $B$. The results on sample $B$ have been shifted 
vertically for better readability.}\label{fig:PAM_strange_large_vs_small}
\end{minipage}
\hspace{0.5cm}
\begin{minipage}[ht]{7.5cm}
\includegraphics[height=6cm,width=7.5cm]{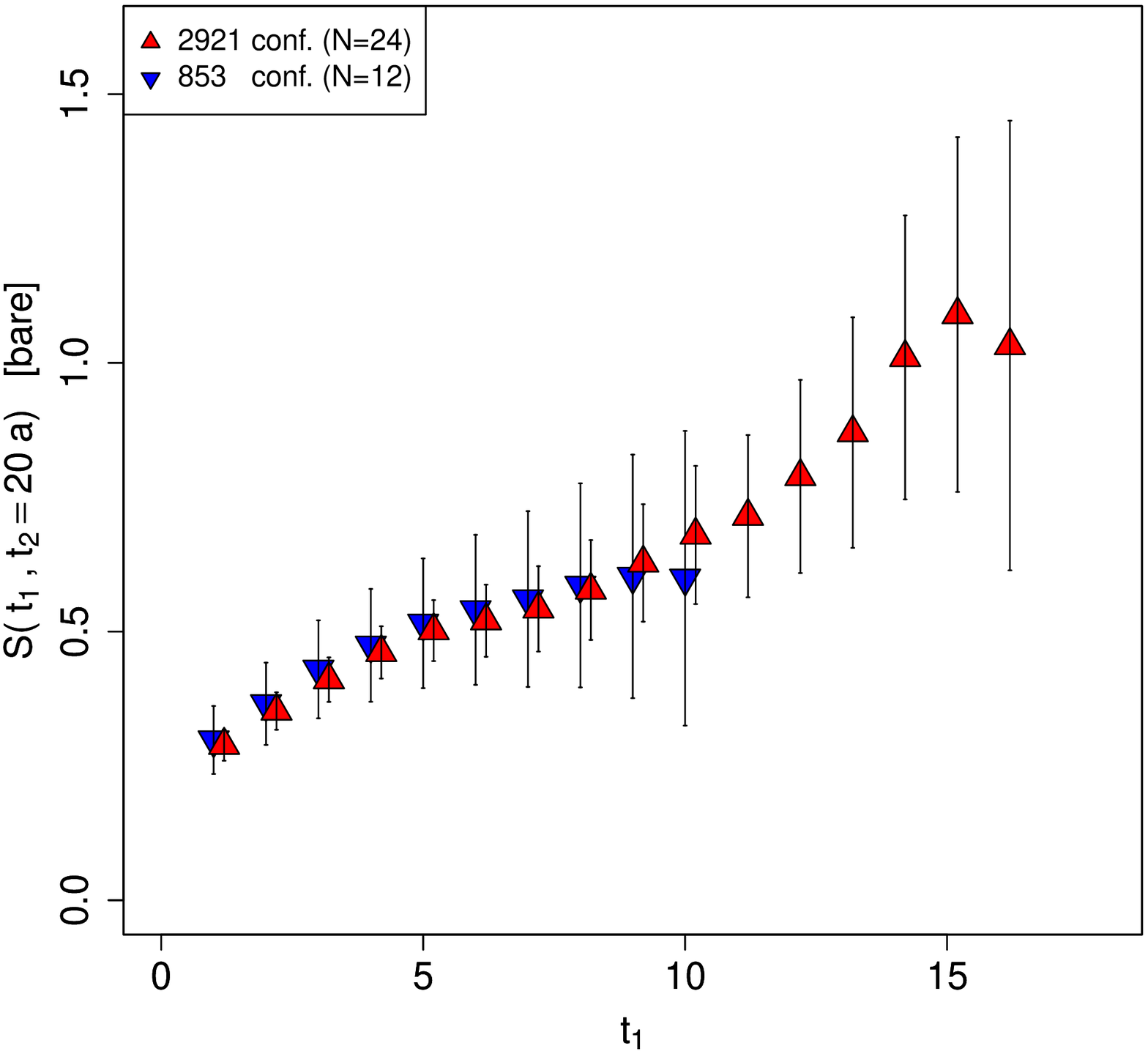}
\caption{Fits of the summed ratio for a fixed value of $t_2=20$ as a 
function of $t_1$ for the two samples $A$ and $B$ .}\label{fig:fits_PAM}
\end{minipage}
\end{figure}

Let us now look at an analysis of the strange quark content using the 
plateau summation method (see \eq{eq:sum_ratio_def}). We show in 
\fig{fig:PAM_strange_large_vs_small} the summed plateau $P_{O_s}$ 
defined in \eq{eq:sum_ratio_def} as a function of the source-sink 
separation for the two samples of measurement $A$ and $B$. For 
better readability we  shifted $P_{O_s}$ vertically in the case of the 
sample of measurements $B$. We recall that for large $t_s$, the bare 
matrix element can be extracted from the slope of $P_{O_s}$. 
We also show using black dotted lines linear fits of $P_{O_s}$ 
choosing the fitting interval $t_s \in [t_1,t_2]$ such that we obtain 
a confidence level larger than $90 \%$. The corresponding 
slopes are  $0.42(8)$ and $0.60(10)$ respectively in the case of 
samples $A$ and $B$. 

In order to study the excited state contaminations we define 
a function $S(t_1,t_2)$ 
as the slope of a linear fit of $P_{O_q}$ in the fitting window 
$[t_1,t_2]$. By construction, for $t_2$ large enough, $S$ should 
be constant as a function of $t_1$ and then will give the bare matrix element 
$\la N | \bar{s}s | N \ra^{(\rm bare)}$. In the following discussion 
we will fix $t_2$ to $t_2=20a$, which appears to be large enough. 
We show in \fig{fig:fits_PAM} $S(t_1,t_2)$ as a function of $t_1$ for 
fixed $t_2=20a$ 
and for the two samples $A$ and $B$.  While the results obtained for 
sample $A$ seem to saturate for $t_1 \gtrsim 6$, the 
results for sample $B$ are constant only for $t_1 \gtrsim 14$. 
Thus we reach a very similar conclusion as discussed above for 
the ratio $R_{O_q}$ of eq.~(\ref{eq:ratio_def}) that a too small 
source-sink separation and insufficient statistics can be 
misleading and that a true plateau has not been reached yet. 
Even with the here employed high statistics to obtain 
a signal at large values of $t_s$, it is not clear  
whether the excited state contributions are negligible. 

Thus, in order to have a convincing evidence that a source-sink separation 
of $t_s=18a \sim 1.5\fm$ is sufficient, it would be necessary to analyze for 
an even larger source-sink while keeping a 
similar statistical accuracy as the one obtained here.  
We stress that with our choice of nucleon  
interpolating fields we do not find a similarly large contribution of excited 
states for other observables, for instance the nucleon axial coupling $g_A$. 

In the light quark sector we observe an effect of excited 
state contamination of the same order of magnitude 
in the corresponding disconnected part. However, here the 
disconnected part contributes 
only about $10 \%$ of the bare matrix element. 
In order to check the size of excited state contributions 
in the connected part, we have analyzed the matrix element
for two source-sink separations, $t_s=12a$ and $t_s=16a$. 
We show in \fig{fig:plateau_light} the results for the bare 
matrix element for these two values of the source-sink separation. 
We observe a change of the plateau value by about 
$\sim 10\%$. Note that even if the effect of excited states is 
thus smaller than in the strange sector, it is  
still large compared to our statistical errors.

\section{Excited states contamination : $y_N$ parameter}

\begin{figure}[h]
\begin{minipage}[ht]{7.5cm}
\vspace{-1.4cm}
\includegraphics[height=6cm,width=7.5cm]{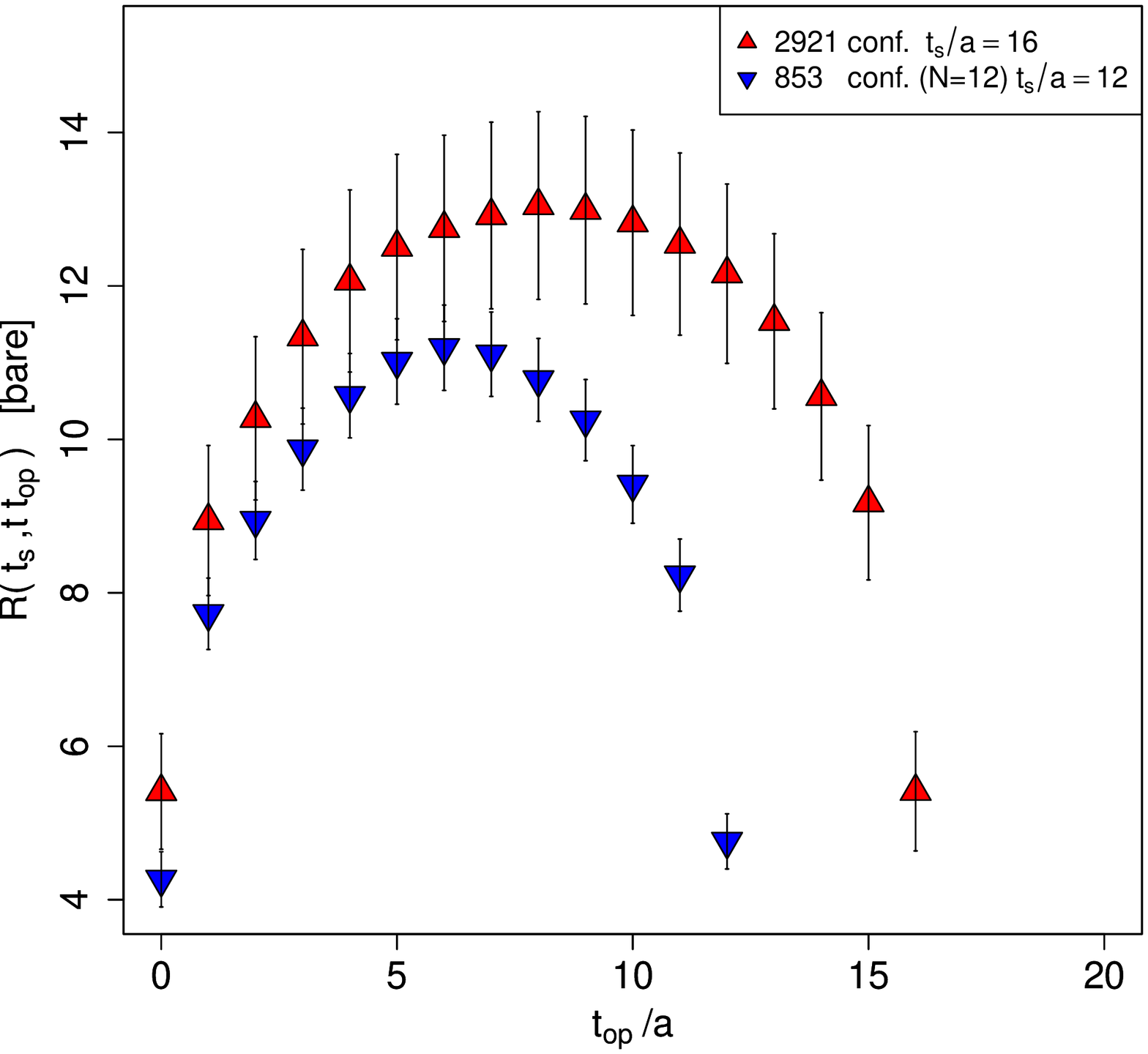}
\caption{The ratio $R_{O_l}(t_s=12a,t_{\rm op})$ and $R_{O_l}(t_s=16a,t_{\rm op})$ for the light 
quark content of the nucleon on samples $A$ and $B$. }\label{fig:plateau_light}
\end{minipage}
\hspace{0.5cm}
\begin{minipage}[ht]{7.5cm}
\includegraphics[height=6cm,width=7.5cm]{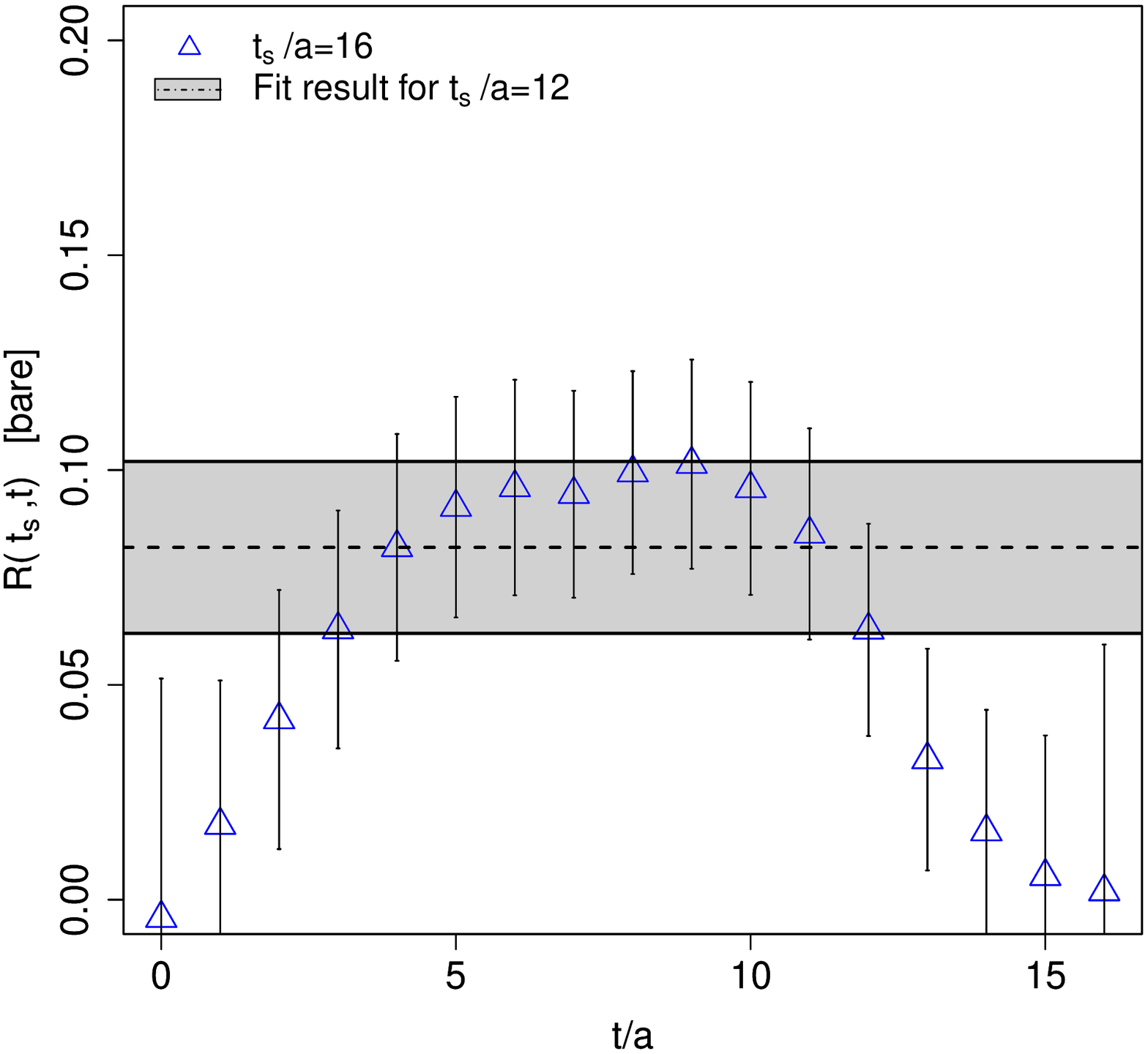}
\caption{Plateau value for the strangeness content of the nucleon 
for two different source-sink separation. The large source-sink separation 
results have been obtained using a larger statistics in the light sector 
(connected and disconnected contributions) and in the strange sector.}\label{fig:strangeness}
\end{minipage}
\end{figure}

Finally, we have also analyzed the contribution of excited states in the determination of 
the $y_N$ parameter. This quantity can be obtained directly by computing the ratio 
of two three-point correlators, see \cite{Dinter:2012tt}.  
As shown in \fig{fig:strangeness}, 
the $y_N$-parameter obtained for a source-sink separation of $t_s=16a$ on the sample $B$  
is compatible with the fit result obtained in \cite{Dinter:2012tt} at a 
source-sink separation of $12a$ with the sample $A$ 
(represented by a grey band). Thus, in the ratio corresponding to 
the $y_N$-parameter a cancellation of systematic effects seems to 
occur and the total systematic error is at most 
of the order of the statistical error, namely $\sim 20\%$.

\section{Conclusion}

As shown in a recent paper \cite{Dinter:2012tt} twisted mass fermions 
at maximal twist 
offer an efficient setup to compute the three point correlators relevant 
for the extraction of the nucleon sigma-terms.
In this proceedings contribution, we have performed a dedicated study of 
excited state contaminations on a single gauge field ensemble corresponding 
to a pion mass of approximately $\approx 390 \mev$. We have shown that 
the strange $\sigma_s$ term is particularly sensitive to excited state contaminations 
that can contribute an about 40\% systematic uncertainty. 
Since the problem is not seen in other nucleon observables, this large 
effect cannot originate from our choice of the nucleon interpolating fields. 
Thus, we suspect that our finding is true in general, also for other
formulations of lattice QCD. 
We therefore conclude that studies of excited state contaminations 
are essential to obtain reliable results. As a consequence of our study, 
we are presently only able to provide 
a lower bound of the systematic error in the direct 
determination of the strange $\sigma_s$ term of size $\gtrsim 40\%$. 
In the light quark sector, the systematic effects is of the 
the order of $\gtrsim  10\%$. We observe a cancellation 
of the effects of excited state contributions in the ratio that determines 
the $y_N$-parameter which lead to a safer determination of this quantity.

In our analysis, we have used 
two approaches, the standard plateau method and the summed ratio method. 
Independently of these two  analysis 
methods, we find it to be necessary to use 
approximately constant statistical errors while the source-sink 
separation is increased to detect the real size of the 
excited state contributions. Such a strategy has, of course, the drawback to 
increase exponentially the number of measurements needed, but it seems to us 
to be a mandatory step to provide reliable values of the scalar quark content of 
the nucleon. 

\section*{Acknowledgments}
This work was performed using HPC resources provided by the JSC Forschungszentrum J\"ulich on the JuGene supercomputer and by GENCI-IDRIS (Grant 2012-052271). It is supported in part by  the DFG Sonder\-for\-schungs\-be\-reich/ Trans\-regio SFB/TR9. Computational resources where partially provided by the Cy-Tera Project NEA Y$\Pi$O$\Delta$OMH/$\Sigma$TPATH/0308/31, which is co-funded by the European
Regional Development Fund and the Republic of Cyprus through the Research Promotion Foundation. A. Vaquero is supported by the Research Promotion Foundation (RPF) of Cyprus under grant $\Pi$PO$\Sigma$E$\Lambda$KY$\Sigma$H/NEO$\Sigma$/0609/16

\end{document}